\documentclass[11pt]{article}
\usepackage{amsmath, amssymb}
\usepackage{graphics,graphicx,color,booktabs}
\usepackage{easybmat}
\usepackage{multirow}
\numberwithin{equation}{section}

\addtocounter{page}{0}
\newtheorem{thm}{Theorem}[section]

\newtheorem{lem}{Lemma}[section]
\allowdisplaybreaks

\begin{document}

\title{\bf Quantile Regression for General Spatial Panel Data Models with Fixed Effects}

\author{Xiaowen Dai$^1$, Zhen Yan$^1$, Maozai Tian$^{1,2}$, ManLai Tang$^3$\\
\small 1. Center for Applied Statistics, School of Statistics, Renmin University of China, Beijing 100872, China \\
\small 2. School of Statistics, Lanzhou University of Finance and Economics Lanzhou, 730101, Gansu, China\\
\small 3. Department of Mathematics and Statistics, Hang Seng Management College, Shatin, Hong Kong, China}
\maketitle

\begin{abstract}
This paper considers the quantile regression model with  both individual fixed effect and
time period effect for general spatial panel data. Instrumental variable quantile regression estimators will be proposed. Asymptotic properties of the proposed estimators will be developed. Simulations are conducted to study the performance of the proposed method. We will illustrate our methodologies using a cigarettes demand data set. 
\end{abstract}

\section{Introduction}

Spatial econometric models have been widely used in many areas (e.g., economics, political science and public health) to deal with spatial interaction effects among geographical units (e.g., jurisdictions, regions, and states).
Recently, the spatial econometrics literature has exhibited a growing interest in the specification and estimation of econometric relationships based on spatial panels, which typically refer to data containing time series observations of a number of spatial units. For instance, Kapoor {\it et al}. (2007) developed a generalized moments (GM) estimator for a space-time model with error components that are both spatially and time-wise correlated. Lee and Yu (2010) proposed the maximum likelihood (ML) estimator for the spatial autoregressive (SAR) panel model with both spatial lag and spatial disturbances. All these works were developed based on  (conditional) mean regression methods. Compared with mean regression methods, the quantile regression (QR) method is more robust and can be adopted to deal with data characterized by different error distributions.

Recently, there has been a growing literature on estimating and testing of QR panel data models. Koenker (2004) introduced a novel approach for the estimation of a QR model for longitudinal data. Galvao {\it et al}. (2011) studied the quantile regression dynamic panel model with fixed effects. Galvao {\it et al}. (2013) investigated the estimation of censored QR models with fixed effects. Galvao {\it et al}. (2015) developed a new minimum distance quantile regression (MD-QR) estimator for panel data models with fixed effects. However,  quantile regression estimation for spatial econometric panel models has not been studied  in existing literature.

This paper focuses on the QR estimations in the general SAR panel data  model with both individual fixed effects and time period specific effects (see, e.g., Lee and Yu, 2010). We employ the instrumental variable quantile regression (IVQR) method to estimate the parameters. The asymptotic properties of the IVQR estimator are also developed. The rest of the paper is organized as follows. Section 2 introduces the SAC panel data model with both individual fixed effects and time period fixed effects, and proposes  the instrumental variable quantile regression (IVQR) estimation procedure. The asymptotic properties of the IVQR estimators are also discussed. Proofs of the theorems in Sections 2 are given in the Appendix. Section 3 reports a simulation study for assessing the finite sample performance of the proposed estimators. An empirical illustration is considered in Section 4. Section 5 concludes the paper.

\section{General spatial autoregressive panel data quantile regression model with both individual and time effects}

 Lee and Yu (2010, Eq. (19)) considered the following general spatial autoregressive panel data model with both individual and time effects
\begin{align} \label{mod:01}
\begin{split}
y_{it}&=\rho\sum_{j\neq i}w_{ij}y_{jt}+X_{it}\beta+\nu_i+\psi_t+u_{it},\quad i=1,\cdots,N,t=1,\cdots,T, \\
u_{it}&=\lambda\sum_{j\neq i}m_{ij}u_{jt}+\varepsilon_{it},
\end{split}
\end{align}
where $y_{it}$ is the dependent variable for subject $i$ at time $t$, $X_{it}$ is a $1\times p$ vector of nonstochastic time varying explanatory variables, $w_{ij}$ is the $(i,j)$th element of the spatial weight matrix $W$ reflecting spatial dependence on $y_{it}$ among cross
sectional units, and $\varepsilon_{it}$ is independent and identically distributed across $i$ and $t$. Similarly, $m_{ij}$ is the $(i,j)$th element of the spatial weight matrix $M$ for the disturbances. The parameters $\nu_i, i=1,\cdots,N$ are fixed effects for the regions while the parameters $\psi_t, t=1,\cdots,T$ are fixed time effects. Interaction effects are reflected in the spatial-temporal lag variable $\displaystyle{\sum_{j\neq i}w_{ij}y_{jt}}$ (and associated scalar parameter $\rho$). In practice, $M$ may or may not be $W$.

The model \eqref{mod:01} can also be written in an alternative form as
\begin{equation} \label{mod:011}
y_{it}=\lambda\sum_{j\neq i}m_{ij}y_{jt}+Z^*_{it}\alpha^*+Z^*_{1i}\nu^*+Z^*_{2t}\psi^*+\varepsilon_{it},
\end{equation}
where $\alpha=(\rho,\beta^T)^T$, $\nu=(\nu_1,\cdots,\nu_N)^T$, $\psi=(\psi_1,\cdots,\psi_T)$, $\alpha^*=(\alpha^T,\lambda\alpha^T)^T$, $\nu^*=(\nu^T,\lambda\nu^T)^T$, $\psi^*=(\psi^T,\lambda\psi^T)^T$,  $Z_1=\mathbf{1}_T\otimes I_N$ is an $NT\times N$ matrix, $Z_2=I_T\otimes\mathbf{1}_N$ is an $NT\times T$ matrix, $\mathbf{1}_J$ is the $J\times1$ vector with all the elements being 1,  $Z^*_{it}=[Z_{it}, -\sum_{j\neq i}m_{ij}Z_{jt}]$, $Z_{it}=[\sum_{j\neq i}w_{ij}y_{jt}, X_{it}]$, $Z^*_{1i}=[Z_{1i}, -\sum_{j\neq i}m_{ij}Z_{1j}]$, $Z^*_{2t}=[Z_{2t}, -\sum_{j\neq i}m_{ij}Z_{2t}]$, $Z_{1i}=h'_{1i}Z_1$ is an indicator variable for the individual effect $\nu_i$, $h_{1i}$ is an $NT\times1$ vector with the $i$th element equal to 1 and the rest equal to 0, $i=1,\cdots,N$, $Z_{2t}=h'_{2t}Z_2$ is an indicator variable for the time effect $\psi_t$, and $h_{2t}$ is an $NT\times1$ vector with the $(t-1)N+1$th element equal to 1 and the rest equal to 0, $t=1,\cdots,T$.

Matrix form of model (\ref{mod:011}) is
\begin{equation} \label{mod:012}
y = \lambda M^{*}y+Z^{*}\alpha^{*}+Z_1^{*}\nu^{*}+Z_2^*\psi^*+\varepsilon,
\end{equation}
where $y=(y_1,\cdots,y_T)^T$ is an $NT\times1$ vector with $y_i=(y_{1i},\cdots,y_{Ni})^T$,  $\varepsilon=(\varepsilon_1,\cdots,\varepsilon_T)^T$ is an $NT\times1$ vector with $\varepsilon_i=(\varepsilon_{1i},\cdots,\varepsilon_{Ni})^T$,  $X=(X_1,\cdots,X_T)^T$ is an $NT\times p$ matrix, $X_i=(X_{1i},\cdots,X_{Ni})^T$, $W^{*}=I_T\otimes W$, $M^{*}=I_T\otimes M$, $W$ and $M$ are both $N\times N$ spatial weight matrices, $Z=[W^*y,X]$, $Z^*=[Z,-M^{*}Z]$, $Z^*_1=[Z_1,-M^{*}Z_1]$, and $Z^*_2=[Z_2,-M^{*}Z_2]$. Here we denote $\theta^*=(\lambda,\alpha^{*T},\nu^{*T},\psi^{*T})^T=\theta^*=(\lambda,\rho,\beta^{*T},\nu^{*T},\psi^{*T})^T$.

We consider the following conditional $\tau$-quantile of response variable:
\begin{equation}
Q_{\tau}(y_{it}|\mathcal{F}_{-it},Z^*_{it},Z^*_{1i},Z^*_{2t})=\lambda(\tau)\sum_{j\neq i}m_{ij}y_{jt}+Z^*_{it}\alpha^*(\tau)+Z^*_{1i}\nu^*(\tau)+Z^*_{2t}\psi^*(\tau),
\end{equation}
where $\tau$ is a quantile in the interval $(0,1)$. We define the objection function by
\begin{equation}
R(\tau,\theta^*)=\sum_{i=1}^{N}\sum_{t=1}^{T}\rho_{\tau}(y_{it}-\lambda\sum_{j\neq i}m_{ij}y_{jt}-Z^*_{it}\alpha^*-Z^*_{1i}\nu^*-Z^*_{2t}\psi^*),
\end{equation}
where $\rho_\tau(u)=u(\tau-I(u\leq0))$ is the check function and $I(\cdot)$ is the indicator function (see, e.g., Koenker, 2005). The estimator $\hat{\theta^*}(\tau)$ can then be obtained by
\begin{equation}
\hat{\theta^*}(\tau)=\underset{\theta^*}{\arg\min}R(\tau,\theta^*).
\end{equation}

\subsection{Instrumental Variable Quantile Regression Estimator (IVQR)} \label{sec:ivqr}

In this section, we employ the instrumental variable quantile regression (IVQR) method for estimation. Let $d_{it}=\sum_{j\neq i}m_{ij}y_{jt}$ denote a scalar endogenous variable, which is related to a vector of instruments $\omega_{it}$. The instruments $\omega_{it}$ are independent of $\varepsilon_{it}$. Consider the objection function for the conditional instrumental quantile relationship:
\begin{equation} \label{ob:iv}
R_{IV}(\tau,\theta^*,\gamma)=\sum_{i=1}^{N}\sum_{t=1}^{T}\rho_{\tau}(y_{it}-\lambda d_{it}-Z^*_{it}\alpha^* -Z^*_{1i}\nu^*-Z^*_{2t}\psi^*-\omega_{it}\gamma).
\end{equation}
Following Chernozhukov and Hansen (2006, 2008) and Galvao (2011), and assuming the availability of instrumental variables $\omega_{it}$, we can derive the IVQR estimator via the following three steps:

Step 1: For a given quantile $\tau$, define a suitable set of values $\{\lambda_j,j=1,\cdots,J;|\lambda|<1\}$. One then minimizes the objective function for $\theta^*,\gamma$ to obtain the ordinary QR estimators
of $\alpha^*,\nu^*,\psi^*,\gamma$:
\begin{equation}
(\hat\alpha^*(\lambda,\tau),\hat\nu^*(\lambda,\tau)),\hat\psi^*(\lambda,\tau)),\hat\gamma(\lambda,\tau))
=\underset{\alpha^*,\nu^*,\psi^*,\gamma}{\arg\min}R_{IV}(\tau,\theta^*,\gamma).
\end{equation}

Step 2: Choose $\hat\lambda(\tau)$ among $\{\lambda_j,j=1,\cdots,J\}$ which makes a weighted distance function defined on $\gamma$ closest to zero:
\begin{equation}
\hat{\lambda}(\tau)=\underset{\lambda\in\mathcal{L}}{\arg\min}\bigg\{\hat\gamma(\lambda,\tau)^T
\hat{A}(\tau)\hat\gamma(\lambda,\tau)\bigg\},
\end{equation}
where $A$ is a positive definite matrix.

Step 3: The estimation of $\alpha^*,\nu^*,\psi^*$ can be obtained, which is respectively $\hat{\alpha}^{*IV}(\hat{\lambda}(\tau),\tau)$, $\hat{\nu}^{*IV}(\hat{\lambda}(\tau),\tau)$, $\hat{\psi}^{*IV}(\hat{\lambda}(\tau),\tau)$.

\subsection{Asymptotic theory}

In this section, we investigate the asymptotic properties of the IVQR estimator in Model \eqref{mod:01}. We impose the following regularity conditions:

\textbf{A1} $\{(y_{it},X_{it})\}$ is independent and identically distributed (i.i.d.) for each fixed $i$ with conditional distribution function $F_{it}$ and differentiable conditional densities, $0<f_{it}<\infty$, with bounded derivatives $f'_{it}$ for $i=1,\cdots,N$ and $t=1,\cdots,T$.

\textbf{A2} For all $\tau\in\mathcal{T}$, $(\lambda(\tau),\alpha^*(\tau))$ is in the interior of the set $\mathcal{L}\times\mathcal{A}$, and $\mathcal{L}\times\mathcal{A}$ is compact and convex.

\textbf{A3} Let $\vartheta=(\theta^{*T},\gamma^T)^T$,
\begin{align}
\Pi(\vartheta,\tau)&=\mathbb{E}[(\tau-I(y<B^{-1}(Z^{*}\alpha^*+Z_1^*\nu^*+Z_2^*\psi^*+E\gamma)))\Delta(\tau)],
\end{align}
 and
 \begin{align}
\Pi(\theta^*,\tau)&=\mathbb{E}[(\tau-I(y<B^{-1}(Z^{*}\alpha^*+Z_1^*\nu^*+Z_2^*\psi^*)))\Delta(\tau)],
\end{align}
where $B=I_{NT}-\lambda M^{*}$ and $\Delta(\tau)=[Z^*,Z_1^*,Z_2^*,E]$. The Jacobian matrices $\frac{\partial\Pi(\theta^*,\tau)}{\partial(\lambda,\alpha^*,\nu^*,\psi^*)}$ and $\frac{\partial\Pi(\vartheta,\tau)}{\partial(\alpha^*,\nu^*,\psi^*,\gamma)}$ are continuous and have full rank uniformly over $\mathcal{A}\times\mathcal{N}\times\mathcal{P}\times\mathcal{G}\times\mathcal{T}$. The parameter space $\mathcal{L}\times\mathcal{A}\times\mathcal{N}\times\mathcal{P}$ is a connected set and the image of $\mathcal{L}\times\mathcal{A}\times\mathcal{N}\times\mathcal{P}$ under the map $\theta^*\mapsto\Pi(\theta^*,\tau)$ is simply connected.

\textbf{A4} Denote $\Omega=\text{diag}(f_{it}(\xi_{it}(\tau)))$, where $\xi_{it}(\tau)=\lambda(\tau)d_{it}+Z^*_{it}\alpha^*(\tau)+Z^*_{1i}\nu^*(\tau)+Z^*_{2t}\psi^*(\tau)
+\omega_{it}\gamma(\tau)$. Let $\tilde{X}=[Z^*,E]$. Then, the following matrices are positive definite:
\begin{align}
\mathbf{J}_{\zeta}&=\underset{N,T\rightarrow\infty}{\lim}\frac{1}{NT}\tilde{X}^{T}M^T_{\tilde{Z}}\Omega M_{\tilde{Z}}\tilde{X}, \\ \mathbf{J}_{\lambda}&=\underset{N,T\rightarrow\infty}{\lim}\frac{1}{NT}\tilde{X}^{T}M^T_{\tilde{Z}}\Omega M_{\tilde{Z}}D,  \\
S&=\underset{N,T\rightarrow\infty}{\lim}\frac{\tau(1-\tau)}{NT}\tilde{X}^TM^T_{\tilde{Z}}M_{\tilde{Z}}\tilde{X}.
\end{align}
where $M_{\tilde{Z}}=I-P_{\tilde{Z}}$ and $P_{\tilde{Z}}=\tilde{Z}(\tilde{Z}^T\Omega\tilde{Z})^{-1}\tilde{Z}^T\Omega$, $\tilde{Z}=[Z_1^*,Z_2^*]$. Let $[\bar{\mathbf{J}}^T_{\alpha^*},\bar{\mathbf{J}}_\gamma^T]$ be a
conformable partition of $\mathbf{J}^{-1}_{\zeta}$ and $H=\bar{\mathbf{J}}^T_\gamma A\bar{\mathbf{J}}_\gamma$. Hence, $\mathbf{J}_{\zeta}$ is invertible and $\mathbf{J}_{\lambda}^{T}H\mathbf{J}_{\lambda}$ is also invertible.

\textbf{A5} $\max\| y_{it}\|=O(\sqrt{NT})$, $\max\| X_{it}\|=O(\sqrt{NT})$ and $\max\|\omega_{it}\|=O(\sqrt{NT})$.

\begin{lem} \label{lem:1}
Denote $\varepsilon_{it}(\tau)=y_{it}-\xi_{it}(\tau)$, and let $\vartheta=(\lambda,\alpha^{*T},\nu^{*T},\psi^{*T},\gamma^T)^T$ be a parameter vector in $\mathcal{V}=\mathcal{L}\times\mathcal{A}\times\mathcal{N}\times\mathcal{P}\times\mathcal{G}$. Let
\begin{equation}
\delta=\begin{pmatrix}
\delta_{\lambda} \\
\delta_{\alpha^*} \\
\delta_{\nu^*} \\
\delta_{\psi^*} \\
\delta_{\gamma}
\end{pmatrix}
=\begin{pmatrix}
\sqrt{NT}(\hat\lambda(\tau)-\lambda(\tau)) \\
\sqrt{NT}(\hat\alpha^*(\tau)-\alpha^*(\tau)) \\
\sqrt{T}(\hat\nu^*-\nu^*) \\
\sqrt{N}(\hat\psi^*-\psi^*) \\
\sqrt{NT}(\hat\gamma(\tau)-\gamma(\tau))
\end{pmatrix}.
\end{equation}
Under conditions A1-A5, we have
\begin{align*}
&\underset{\vartheta\in\mathcal{V}}
{\sup}\frac{1}{NT}\bigg|\sum_{i=1}^{N}\sum_{t=1}^{T}\bigg[\rho_{\tau}\bigg(\varepsilon_{it}(\tau)-\frac{d_{it}\delta_{\lambda}}{\sqrt{NT}}
-\frac{Z^*_{it}\delta_{\alpha^*}}{\sqrt{NT}}-\frac{Z^*_{1i}\delta_{\nu^*}}{\sqrt{T}}-\frac{Z^*_{2t}\delta_{\psi^*}}
{\sqrt{N}}-\frac{\omega_{it}\delta_{\gamma}}{\sqrt{NT}}\bigg)-\rho_{\tau}(\varepsilon_{it}(\tau))\\
&-E\bigg[\rho_{\tau}\bigg(\varepsilon_{it}(\tau)-\frac{d_{it}\delta_{\lambda}}{\sqrt{NT}}
-\frac{Z^*_{it}\delta_{\alpha^*}}{\sqrt{NT}}-\frac{Z^*_{1i}\delta_{\nu^*}}
{\sqrt{T}}-\frac{Z^*_{2t}\delta_{\psi^*}}{\sqrt{N}}-\frac{\omega_{it}\delta_{\gamma}}{\sqrt{NT}}\bigg)
-\rho_{\tau}(\varepsilon_{it}(\tau))\bigg]\bigg]\bigg|=o_p(1).
\end{align*}
\end{lem}

\begin{thm}[Consistency] \label{th:1}
Under conditions A1-A5, $(\lambda(\tau),\alpha^*(\tau),\nu^*,\psi^*)$ uniquely solves the equation $\mathbb{E}[(\tau-I(y<B^{-1}(Z^{*}\alpha^*+Z_1^*\nu^*+Z_2^*\psi^*)))\Delta(\tau)]=0$ over $\mathcal{L}\times\mathcal{A}\times\mathcal{N}\times\mathcal{P}$ and $(\lambda(\tau),\alpha^*(\tau),\nu^*,\psi^*)$ is consistently estimable. Therefore, the parameters $\rho,\beta,\nu,\psi$ are also consistently estimable.
\end{thm}

\begin{thm}[Asymptotic distribution] \label{th:2}
Under conditions A1-A5 and Lemma \ref{lem:1}, for a given $\tau\in(0,1)$, $\hat\theta=(\hat\lambda,\hat\alpha^*)=(\hat\lambda,\hat\rho,\hat\beta,\widehat{\lambda\rho},\widehat{\lambda\beta})$ converges to a Gaussian distribution:
\begin{equation}
\sqrt{NT}(\hat\theta(\tau)-\theta(\tau))\stackrel{d}{\rightarrow}N(0,\Lambda),
\end{equation}
where $\Lambda=J^{T}SJ$, $S=\underset{N,T\rightarrow\infty}{\lim}\frac{\tau(1-\tau)}{NT}\tilde{X}^TM^T_{\tilde{Z}}M_{\tilde{Z}}\tilde{X}$,
$\tilde{X}=[Z^*,E]$, $J=(K^T,L^T)$, $M_{\tilde{Z}}=I-P_{\tilde{Z}}$, $P_{\tilde{Z}}=\tilde{Z}(\tilde{Z}^T\Omega\tilde{Z})^{-1}\tilde{Z}^T\Omega$, $\tilde{Z}=[Z_1^*,Z_2^*]$, $\Omega=diag(f_{it}(\xi_{it}(\tau)))$, $\mathbf{J}_{\lambda}=\underset{N,T\rightarrow\infty}{\lim}\frac{1}{NT}\tilde{X}^{T}M^T_{\tilde{Z}}\Omega M_{\tilde{Z}}D$, $L=\bar{J}_{\alpha^*}M$, $M=I-\mathbf{J}_{\lambda}K$, $\mathbf{J}_{\zeta}=\underset{N,T\rightarrow\infty}{\lim}\frac{1}{NT}\tilde{X}^{T}M^T_{\tilde{Z}}\Omega M_{\tilde{Z}}\tilde{X}$, $K=(\mathbf{J}_{\lambda}^{T}H\mathbf{J}_{\lambda})^{-1}
\mathbf{J}_{\lambda}^{T}H$, $H=\bar{\mathbf{J}}^T_\gamma A\bar{\mathbf{J}}_\gamma$ and
$[\bar{\mathbf{J}}^T_{\alpha^*},\bar{\mathbf{J}}_\gamma^T]$ is a
conformable partition of $\mathbf{J}^{-1}_{\zeta}$.
\end{thm}

\section{Monte Carlo simulations}

In this section, we report the results of a Monte Carlo study in which we assess the finite sample performance of the IVQR estimators proposed in Section 2. For comparison purpose, we generate the samples being considered in the design of Lee anf Yu (2010):
\begin{align*}
y&=\rho W^{*}y+X\beta+Z_1\nu+Z_2\psi+u, \\
u&=\lambda M^{*}u+\varepsilon,
\end{align*}
where $\theta_{0}=(\rho_0, \lambda_0, \beta_0)^T$, $\theta_{0}^{a}=(0.2, 0.5, 1)^T$ and $\theta_{0}^{b}=(0.5, 0.2, 1)^T$. Here, $X$, $\nu$, $\psi$ are drawn independently from $N(0,1)$ and both the spatial weights matrices $W$ and $M$ are the same rook matrices. We use some combinations of $T=5, 10$, and $n=49$. For the disturbance errors, we consider the standard normal (i.e., N(0, 1)) and Cauchy (i.e., $t_1$) distributions.

For each set of generated sample observations, we calculate the IVQR estimators. This step is repeated for 2000 times. We consider the bias and root mean squared error (RMSE) for the MLE, QMLE, OLS and IVQR. The quantile regression based estimators are calculated for quantiles $\tau=(0.25,0.5,0.75)$. For the IVQR estimator, we employed $y_{it-1}$ as instrument. The results are summarized in Table 1-4.

Table 1-4 show that the IVQR estimator performs better than the other estimators in $t_1$ settings. In general, we find that the biases and RMSEs associated with $\tau=0.5$  are slightly smaller for the  IVQR estimator. Under normal disturbance errors, the IVQR estimators for $\lambda$ performs better than the other estimators while those for $\rho$ and $\beta$ have similar biases and RMSEs as the OLS estimators, but a bit larger than the MLE and QMLE estimators. For Cauchy disturbance errors, our proposed IVQR estimators outperform the other estimators as we do not impose any finite moment assumption on the distrubance errors. Therefore, we conclude that the proposed IVQR is more robust in practice.

\begin{center}
\begin{table} \small
\caption{Bias and RMSE of various estimators (with both individual and time effects) with $N = 49$ and $T = 5$. }
\begin{tabular}{ccccccccc} \toprule
\multirow{2}*{Distribution}&\multirow{2}*{\quad}&\multirow{2}*{\quad}&\multicolumn{3}{c}{IVQR}
&\multirow{2}*{MLE}&\multirow{2}*{QMLE} &\multirow{2}*{OLS} \\ \cline{4-6}
                      &                    &                    &$\tau=0.25$&$\tau=0.50$&$\tau=0.75$&    &  &\\ \hline
\multirow{12}*{$N(0,1)$}&\multirow{2}*{$\rho=0.2$}    &Bias &0.1278&0.1256&0.1278&0.0271&0.0121 &0.1164 \\
                       &                             &RMSE &0.2293&0.2176&0.2286&0.1243&0.1408  &0.1935\\
                       &\multirow{2}*{$\lambda=0.5$} &Bias &0.0085&0.0056&0.0075&-0.0904&-0.0300  &0.1877 \\
                       &                             &RMSE &0.0408&0.0406&0.0410&0.1618&0.1558  &0.2382\\
                       &\multirow{2}*{$\beta=1$}     &Bias &0.0096&0.0037&0.0049&0.0021&-0.0020  &-0.0276\\
                       &                              &RMSE &0.0975&0.0872&0.0976&0.0749&0.0764  &0.0786\\ \cline{2-9}
                       &\multirow{2}*{$\rho=0.5$}    &Bias&0.2382&0.2381&0.2472&-0.0382&-0.0167 &0.2264 \\
                        &                               &RMSE&0.2844&0.2749&0.2887&0.1129&0.1238  &0.2599 \\
                        &\multirow{2}*{$\lambda=0.2$} &Bias&-0.0001&0.0001&0.0001&0.0183&0.0017 &-0.0237 \\
                         &                              &RMSE &0.0041&0.0040&0.0041&0.1455&0.1658 &0.2440 \\
                        &\multirow{2}*{$\beta=1$}       &Bias &-0.0117&-0.0057&-0.0105&-0.0017&-0.0042 &-0.0299 \\
                         &                              &RMSE&0.1063&0.0926&0.1065&0.0733&0.0738 &0.0817 \\ \hline
\multirow{12}*{$t_1$}&\multirow{2}*{$\rho=0.2$}    &Bias &0.0199&0.0136&0.0183&0.0512&0.0528&0.2183 \\
                       &                             &RMSE &0.0579&0.0402&0.0598&0.1815&0.1775&0.2760    \\
                       &\multirow{2}*{$\lambda=0.5$} &Bias &0.0029&0.0022&0.0043&-0.1057&-0.1095&0.2114  \\
                       &                             &RMSE &0.0410&0.0404&0.0412&0.1937&0.1917&0.2608  \\
                       &\multirow{2}*{$\beta=1$}     &Bias &0.0081&0.0033&-0.0007&-1.3403&-5.6184&-0.5514  \\
                       &                              &RMSE &0.2253&0.1513&0.2243&57.2094&153.2562&47.4065  \\ \cline{2-9}
                       &\multirow{2}*{$\rho=0.5$}    &Bias  &0.0454&0.0311&0.0484&-0.1077&-0.1031&0.4463  \\
                        &                               &RMSE  &0.0728&0.0502&0.0759&0.1841&0.1823&0.4645  \\
                        &\multirow{2}*{$\lambda=0.2$} &Bias  &0.0031&0.0021&0.0030&0.0546&0.0551&-0.0416  \\
                         &                              &RMSE  &0.0041&0.0041&0.0041&0.1763&0.1740&0.1947  \\
                        &\multirow{2}*{$\beta=1$}       &Bias  &0.0098&0.0023&0.0183&-0.2703&0.8199&-0.2556  \\
                         &                              &RMSE&0.2127&0.1521&0.2087&21.7709&17.4982&14.4871  \\ \bottomrule
\end{tabular}
\end{table}
\end{center}

\begin{center}
\begin{table}\small
\caption{Bias and RMSE of various  estimators (with both individual and time effects) \
with $N = 49$ and $T = 10$. }
\begin{tabular}{ccccccccc} \toprule
\multirow{2}*{Distribution}&\multirow{2}*{\quad}&\multirow{2}*{\quad}&\multicolumn{3}{c}{IVQR} &\multirow{2}*{MLE}&\multirow{2}*{QMLE}&\multirow{2}*{OLS}  \\ \cline{4-6}
                      &                    &                    &$\tau=0.25$&$\tau=0.50$&$\tau=0.75$ & & & \\ \hline
\multirow{12}*{$N(0,1)$}&\multirow{2}*{$\rho=0.2$}    &Bias &0.1267&0.1248&0.1274&0.0241&0.0056 &0.1070 \\
                       &                             &RMSE &0.1800&0.1753&0.1789&0.0889&0.0988 &0.1462 \\
                       &\multirow{2}*{$\lambda=0.5$} &Bias &0.0007&0.0009&0.0010&-0.0779&-0.0137 &0.2109  \\
                       &                             &RMSE &0.0406&0.0410&0.0403&0.1198&0.1040 &0.2312 \\
                       &\multirow{2}*{$\beta=1$}     &Bias &0.0012&0.0001&0.0013&0.0038&-0.0001 &-0.0323 \\
                       &                              &RMSE &0.0660&0.0616&0.0666&0.0489&0.0500 &0.0598 \\ \cline{2-9}
                       &\multirow{2}*{$\rho=0.5$}    &Bias &0.2326&0.2246&0.2300&-0.0305&-0.0064  &0.2328\\
                        &                               &RMSE &0.2576&0.2489&0.2549&0.0794&0.0839 &0.2491 \\
                        &\multirow{2}*{$\lambda=0.2$} &Bias &0.0002&0.0000&0.0001&-0.0178&-0.0005 &0.0116  \\
                         &                              &RMSE &0.0041&0.0040&0.0041&0.0996&0.1126  &0.1471 \\
                        &\multirow{2}*{$\beta=1$}       &Bias &-0.0301&-0.0268&-0.0295&0.0001&-0.0013 &-0.0293 \\
                         &                              &RMSE&0.0717&0.0689&0.0736&0.0471&0.0471 &0.0577  \\ \hline
\multirow{12}*{$t_1$}&\multirow{2}*{$\rho=0.2$}    &Bias &0.0131&0.0064&0.0105&0.0571&0.0613&0.2211  \\
                       &                             &RMSE  &0.0397&0.0180&0.0344&0.1639&0.1615&0.2533   \\
                       &\multirow{2}*{$\lambda=0.5$} &Bias  &0.0016&0.0003&0.0030&-0.1065&-0.1093&0.2330  \\
                       &                             &RMSE  &0.0406&0.0408&0.0409&0.1781&0.1775&0.2565  \\
                       &\multirow{2}*{$\beta=1$}     &Bias &-0.0034&0.0001&-0.0008&4.5772&2.9534&-2.0496  \\
                       &                              &RMSE  &0.1365&0.0942&0.1316&151.1233&132.0682&48.2493  \\ \cline{2-9}
                       &\multirow{2}*{$\rho=0.5$}    &Bias  &0.0205&0.0132&0.0211&-0.1010&-0.1017&0.4500  \\
                        &                               &RMSE &0.0373&0.0242&0.0398&0.1696&0.1723&0.4606  \\
                        &\multirow{2}*{$\lambda=0.2$} &Bias  &0.0001&0.0001&-0.0001&0.0533&0.0527&-0.0145  \\
                         &                              &RMSE  &0.0040&0.0040&0.0041&0.1617&0.1609&0.1331  \\
                        &\multirow{2}*{$\beta=1$}       &Bias &-0.0053&-0.0048&-0.0066&-0.4902&-2.9283&0.4147  \\
                         &                              &RMSE &0.1322&0.0947&0.1308&15.7727&82.8778&15.9012  \\ \bottomrule
\end{tabular}
\end{table}
\end{center}

\begin{center}
\begin{table}\small
\caption{Bias and RMSE of various estimators (with individual effect only) with $N = 49$ and $T = 5$. }
\begin{tabular}{cccccccc} \toprule
\multirow{2}*{Distribution}&\multirow{2}*{\quad}&\multirow{2}*{\quad}&\multicolumn{3}{c}{IVQR}
&\multirow{2}*{MLE(QMLE)} &\multirow{2}*{OLS} \\ \cline{4-6}
                      &                    &                    &$\tau=0.25$&$\tau=0.50$&$\tau=0.75$&      &\\ \hline
\multirow{12}*{$N(0,1)$}&\multirow{2}*{$\rho=0.2$}    &Bias &0.0969&0.0935&0.0961&0.0096 &0.0932 \\
                       &                             &RMSE &0.2088&0.1989&0.2127&0.1380  &0.1756\\
                       &\multirow{2}*{$\lambda=0.5$} &Bias &-0.0028&-0.0010&0.0010&-0.0279  &0.2372 \\
                       &                             &RMSE &0.0412&0.0409&0.0409&0.1485  &0.2629\\
                       &\multirow{2}*{$\beta=1$}     &Bias &-0.0092&-0.0076&-0.0104&-0.0027  &-0.0372\\
                       &                              &RMSE &0.1004&0.0878&0.0973&0.0766  &0.0804\\ \cline{2-8}
                       &\multirow{2}*{$\rho=0.5$}    &Bias &0.2121&0.2063&0.2176&-0.0173 &0.2125 \\
                        &                               &RMSE &0.2694&0.2552&0.2737&0.1163  &0.2480 \\
                        &\multirow{2}*{$\lambda=0.2$} &Bias &0.0001&0.0000&0.0000&0.0021 &0.0510 \\
                         &                              &RMSE &0.0041&0.0041&0.0041&0.1590 &0.2172 \\
                        &\multirow{2}*{$\beta=1$}       &Bias &-0.0366&-0.0306&-0.0370&-0.0039 &-0.0340 \\
                         &                              &RMSE&0.1078&0.0942&0.1077&0.0737 &0.0820 \\ \hline
\multirow{12}*{$t_1$}&\multirow{2}*{$\rho=0.2$}    &Bias &0.0164&0.0088&0.0146&-0.4973&0.1838 \\
                       &                             &RMSE &0.0555&0.0362&0.0523&0.5068&0.2449   \\
                       &\multirow{2}*{$\lambda=0.5$} &Bias &0.0018&-0.0005&-0.0002&0.4453&0.2798 \\
                       &                             &RMSE &0.0410&0.0408&0.0410&0.4489&0.3053 \\
                       &\multirow{2}*{$\beta=1$}     &Bias &0.0005&0.0019&-0.0090&-0.2086&0.2149 \\
                       &                              &RMSE &0.2122&0.1555&0.2172&6.7370&20.5035 \\ \cline{2-8}
                       &\multirow{2}*{$\rho=0.5$}    &Bias  &0.0349&0.0231&0.0367&-0.7019&0.4122 \\
                        &                               &RMSE  &0.0654&0.0450&0.0777&0.7774&0.4327 \\
                        &\multirow{2}*{$\lambda=0.2$} &Bias  &0.0001&0.0001&-0.0001&0.6524&-0.0024 \\
                         &                              &RMSE  &0.0042&0.0041&0.0041&0.7311&0.1918 \\
                        &\multirow{2}*{$\beta=1$}       &Bias  &-0.0050&-0.0056&-0.0021&-0.8846&-0.1423 \\
                         &                              &RMSE&0.2011&0.1484&0.2052&17.5186&11.0912 \\ \bottomrule
\end{tabular}
\end{table}
\end{center}

\begin{center}
\begin{table}\small
\caption{Bias and RMSE of various estimators (with individual effects only) with $N = 49$ and $T = 10$. }
\begin{tabular}{cccccccc} \toprule
\multirow{2}*{Distribution}&\multirow{2}*{\quad}&\multirow{2}*{\quad}&\multicolumn{3}{c}{IVQR} &\multirow{2}*{MLE(QMLE)}&\multirow{2}*{OLS}  \\ \cline{4-6}
                      &                    &                    &$\tau=0.25$&$\tau=0.50$&$\tau=0.75$ &  & \\ \hline
\multirow{12}*{$N(0,1)$}&\multirow{2}*{$\rho=0.2$}    &Bias &0.0955&0.0920&0.0960&0.0040 &0.0946 \\
                       &                             &RMSE &0.1570&0.1523&0.1586&0.0949 &0.1358 \\
                       &\multirow{2}*{$\lambda=0.5$} &Bias &-0.0014&0.0002&0.0004&-0.0110 &0.2602  \\
                       &                             &RMSE &0.0409&0.0407&0.0414&0.0945 &0.2698 \\
                       &\multirow{2}*{$\beta=1$}     &Bias &-0.0047&-0.0055&-0.0022&-0.0005 &-0.0364 \\
                       &                              &RMSE &0.0647&0.0588&0.0630&0.0492 &0.0602 \\ \cline{2-8}
                       &\multirow{2}*{$\rho=0.5$}    &Bias &0.2123&0.2130&0.2115&-0.0066  &0.2118\\
                        &                               &RMSE &0.2379&0.2370&0.2361&0.0762 &0.2283 \\
                        &\multirow{2}*{$\lambda=0.2$} &Bias &0.0000&0.0001&0.0000&0.0007 &0.0755  \\
                         &                              &RMSE &0.0041&0.0041&0.0041&0.1053  &0.1636 \\
                        &\multirow{2}*{$\beta=1$}       &Bias &-0.0335&-0.0316&-0.0338&-0.0011 &-0.0362 \\
                         &                              &RMSE&0.0713&0.0718&0.0716&0.0466 &0.0621  \\ \hline
\multirow{12}*{$t_1$}&\multirow{2}*{$\rho=0.2$}    &Bias &0.0078&0.0050&0.0082&-0.3896&0.1907 \\
                       &                             &RMSE  &0.0253&0.0164&0.0249&0.4027&0.2215  \\
                       &\multirow{2}*{$\lambda=0.5$} &Bias  &0.0005&0.0015&-0.0001&0.3629&0.2919 \\
                       &                             &RMSE  &0.0411&0.0410&0.0405&0.3696&0.3064 \\
                       &\multirow{2}*{$\beta=1$}     &Bias &0.0008&0.0003&-0.0007&-1.4438&-2.3930 \\
                       &                              &RMSE  &0.1358&0.0952&0.1349&22.4584&34.5131 \\ \cline{2-8}
                       &\multirow{2}*{$\rho=0.5$}    &Bias  &0.0202&0.0105&0.0194&-0.0082&0.4118 \\
                        &                               &RMSE &0.0363&0.0204&0.0351&0.4977&0.4207 \\
                        &\multirow{2}*{$\lambda=0.2$} &Bias  &-0.0001&0.0001&-0.0001&-0.0192&0.0218 \\
                         &                              &RMSE  &0.0041&0.0040&0.0041&0.5188&0.1456 \\
                        &\multirow{2}*{$\beta=1$}       &Bias &-0.0066&-0.0030&-0.0067&0.5884&-1.0812 \\
                         &                              &RMSE &0.1263&0.0904&0.1274&25.4738&47.3648 \\ \bottomrule
\end{tabular}
\end{table}
\end{center}

Remark: Lee and Yu (2010) demonstrated the transformation approach (QMLE) and the direct approach (MLE) yield the same estimate of $(\rho,\lambda,\beta)$ when considering the individual effects only.

\section{Illustration}

In this section, we use the cigarette demand data set to illustrate our methodologies. The data set is based on a panel of 46 states over 30 time
periods (1963-1992), which has been analyzed by many authors (see, Baltagi and Levin, 1992; Baltagi,
2001; Baltagi, Griffin, and Xiong, 2000; Yang, 2006; Elhorst, 2005; Kelejian and Piras, 2013). The QR model with
both individual and time-period effects is given by:
\begin{align} \label{mod:eg1}
\begin{split}
Q_{\tau}(\log C_{it}|\mathcal{F}_{-it},Z^*_{it},Z^*_{1i},Z^*_{2t})=&\lambda(\tau)\sum_{j\neq i}m_{ij}\log C_{jt}+Z^*_{it}\alpha^*(\tau)+Z^*_{1i}\nu^*(\tau)+Z^*_{2t}\psi^*(\tau),\\
& i=1, \cdots, 46, t=1, \cdots, 30,
\end{split}
\end{align}
where $C_{it}$ is real per capita sales of cigarettes by persons of smoking age (14 years
and older), $Z^*_{it}=[Z_{it},-\sum_{j\neq i}m_{ij}Z_{jt}]$, $Z_{it}=[\sum_{j\neq i}w_{ij}\log C_{jt},X_{it}]$, $X_{it}=[\log P_{it},\log Y_{it}]$, $P_{it}$ is the average retail price of a pack of cigarettes measured in real terms, and $Y_{it}$ is real per capita disposable
income. Here, we choose $\log C_{it-1}$ as instruments.

We estimate the parameters using the IVQR, MLE, and OLS methods. The results are presented in Table 5.
The first three columns are  the IVQR estimates for $\tau = 0.25, 0.50, 0.75$, and the last two columns
correspond to the MLE and OLS estimates respectively.
The top half of table presents the estimations with both the individual and time-period effects while the bottom half of table shows the estimations  with individual effects only.
We can see that the IVQR estimates at different quantiles (i.e., $\tau = 0.25, 0.50, 0.75$) are quite different from the MLE and OLS estimates. In particular, the signs of the estimates for $\rho$ and $\lambda$ are different among IVQR, MLE and OLS methods. Besides, the sign of the MLE estimates for $\rho$ and $\lambda$  change when the individual effect is omitted from the analysis.

\begin{center}
\begin{table}
\caption{Estimation Results of Cigarette Demand Using general spatial panel data models.}
\begin{tabular}{cccccc} \toprule
\multirow{2}*{Parameter}&\multicolumn{3}{c}{IVQR}           &\multirow{2}*{MLE} &\multirow{2}*{OLS}  \\ \cline{2-4}
                        &$\tau=0.25$&$\tau=0.50$&$\tau=0.75$&                   &    \\ \hline
\multicolumn{6}{c}{\textbf{With both individual and time-period effects}} \\
$\rho$                  &-0.5249    &-0.7183    &-0.3087    &0.3547             &-0.1747\\
$\lambda$               &0.1900     &0.4800     &0.0400     &-0.3289            &-0.1747   \\
log average cigarettes retail price&-0.4108    &-0.4284    &-0.4552    &-1.0886            &-0.6115\\
log disposable income   &0.2652     &0.3308     &0.3320     &0.4706             &0.3955\\
\multicolumn{6}{c}{\textbf{With individual effects only}} \\
$\rho$                  &-0.4617    &-0.9111    &-1.1878    &0.3938             &-0.1191\\
$\lambda$               &0.1800     &0.7200     &0.9900     &0.6168            &-0.1191   \\
log average cigarettes retail price&-0.3699    &-0.3699    &-0.4173    &-3.2356            &-0.5464\\
log disposable income   &0.3073     &0.3446     &0.3713     &0.5310             &0.4370
\\ \bottomrule
\end{tabular}
\end{table}
\end{center}

Figure 1 presents a complete analysis, which considers other quantiles of the conditional cigarettes demand distribution.
Similarly, the top panel presents the estimations which are with both individual and time-period effects and the bottom panel shows the estimations which are with individual effects only.
The $x$-axis presents the quantiles and $y$-axis presents the estimations of  parameters (red lines) and their corresponding confidence intervals (blue lines).
We find that the cigarettes retail price has negative effect to the capita sales of cigarettes and disposable income has positive effect to the capita sales of cigarettes.
In the presence of both individual and time-period effects, the estimates of capita sales of cigarettes are larger at extreme quantiles than those at other quantiles. The estimates of disposable income become larger along with the higher quantiles.
However, in the presence of individual effect only, the estimates of parameters are larger at the middle quantiles.

\begin{figure}[htb]
\begin{center}
\scalebox{0.8}[0.8]{\includegraphics{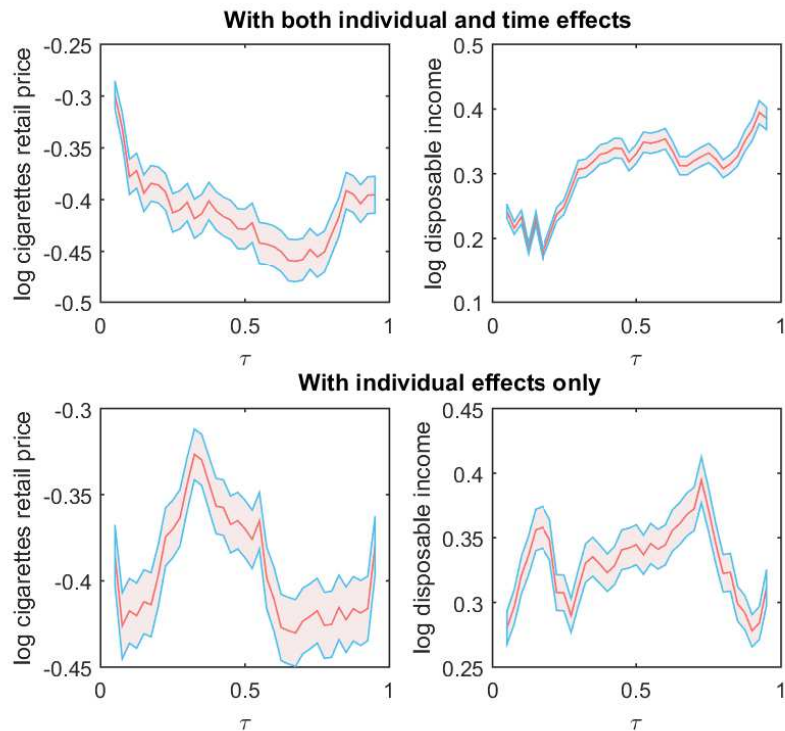}}
\end{center}
\caption{(a)-(b) Quantile effects of the log average retail price of a pack of cigarettes and the log disposable income with both individual and time-period effects. (c)-(d) Quantile effects of the log average retail price of a pack of cigarettes and the log disposable income with individual effects only. The areas represent 90\% point-wise confidence intervals. }
\end{figure}

\section{Conclusion}

In this paper, we investigate the instrumental variable quantile (IVQR) estimation of general spatial autoregressive
panel data model with fixed effects. The model with both individual and time-period effects is considered. The asymptotic properties are studied.
Monte Carlo results are provided to show that the proposed methodology is robust to  error distributions with un-defined moments.

\section{Acknowledgements}

The work was partially supported by National Natural Science Foundation of China (No.11271368), Project supported by the Major Program of Beijing Philosophy and Social Science Foundation of China (No. 15ZDA17), Project of Ministry of Education supported by the Specialized Research Fund for the Doctoral Program of Higher Education of China (Grant No. 20130004110007), The Key Program of National Philosophy and Social Science Foundation Grant (No. 13AZD064), The Fundamental Research Funds for the Central Universities, and the Research Funds of Renmin University of China (No. 15XNL008), and The Project of Flying Apsaras Scholar of Lanzhou University of Finance \& Economics.

\section*{Appendix: Proofs}

Proof of Lemma \ref{lem:1} is similar to that of Lemma 2 in Galvao (2011) and is hence omitted here.

\subsection*{1\quad Proof of Theorem \ref{th:1}}

\textbf{Proof.} Firstly, following Chernozhukov and Hansen (2006), $(\lambda(\tau), \alpha^*(\tau), \nu^*,\psi^*)$ uniquely solves the problem for each $\tau$.

To prove the consistency of the parameter, we need to show that under conditions A1-A5, $\hat\theta^*(\tau)=\theta^*(\tau)+o_p(1)$. Let
$$
\mathcal{P}: \vartheta\mapsto\rho_\tau(y-\lambda D-Z^{*}\alpha^*-Z_1^*\nu^*-Z_2^*\psi^*),
$$
and $\mathcal{P}$ is continuous. Under condition Lemma \ref{lem:1}, we have that $\|\hat\vartheta^*(\lambda,\tau)-\vartheta^*(\lambda,\tau)\|\stackrel{P}{\rightarrow}0$ for $\vartheta^*=(\lambda,\alpha^*,\nu^*,\psi^*,\gamma)$, which implies that $\|\|\hat\gamma(\lambda,\tau)\|-\|\gamma(\lambda,\tau)\|\|\stackrel{P}{\rightarrow}0$. By Corollary 3.2.3 in van der Vaart and Wellner (1996), we have $\|\hat\lambda(\tau)-\lambda(\tau)\|\stackrel{P}{\rightarrow}0$. Therefore, $\|\hat\alpha^*(\hat\lambda(\tau),\tau)-\alpha^*(\tau)\|\stackrel{P}{\rightarrow}0$, $\|\hat\nu^*(\hat\lambda(\tau),\tau)-\nu^*\|\stackrel{P}{\rightarrow}0$, $\|\hat\psi^*(\hat\lambda(\tau),\tau)-\psi^*\|\stackrel{P}{\rightarrow}0$,
and $\|\hat\gamma(\hat\lambda(\tau),\tau)-0\|\stackrel{P}{\rightarrow}0$. Hence, $\|\hat\theta^*(\tau)-\theta^*(\tau)\|\stackrel{P}{\rightarrow}0$ and the theorem follows. $\square$

\subsection*{2\quad Proof of Theorem \ref{th:2}}

For any $\hat\lambda(\tau)\stackrel{P}{\rightarrow}\lambda(\tau)(\delta_{\lambda}\stackrel{P}{\rightarrow}0)$, we
can write the objective function defined in equation \eqref{ob:iv} as
$$
V_{IV}(\alpha)=\sum_{i=1}^{N}\sum_{t=1}^{T}\bigg[\rho_{\tau}\bigg(\varepsilon_{it}(\tau)-\frac{d_{it}\delta_\lambda}{\sqrt{NT}}
-\frac{Z^*_{it}\delta_{\alpha^*}}{\sqrt{NT}}-\frac{Z^*_{1i}\delta_{\nu^*}}{\sqrt{T}}-\frac{Z^*_{2t}\delta_{\psi^*}}
{\sqrt{N}}-\frac{\omega_{it}\delta_{\gamma}}{\sqrt{NT}}\bigg)-\rho_{\tau}(\varepsilon_{it}(\tau)\bigg]
$$
where $\varepsilon_{it}(\tau)=y_{it}-\xi_{it}(\tau)$, $\xi_{it}(\tau)=\lambda(\tau)d_{it}+Z^*_{it}\alpha^*(\tau)+Z^*_{1i}\nu^*(\tau)+Z^*_{2t}\psi^*(\tau)
+\omega_{it}\gamma(\tau)$, and
$$
\delta=\begin{pmatrix}
\delta_{\lambda} \\
\delta_{\alpha^*} \\
\delta_{\nu^*} \\
\delta_{\psi^*} \\
\delta_{\gamma}
\end{pmatrix}
=\begin{pmatrix}
\sqrt{NT}(\hat\lambda(\tau)-\lambda(\tau)) \\
\sqrt{NT}(\hat\alpha^*(\tau)-\alpha^*(\tau)) \\
\sqrt{T}(\hat\nu^*-\nu^*) \\
\sqrt{N}(\hat\psi^*-\psi^*) \\
\sqrt{NT}(\hat\gamma(\tau)-\gamma(\tau))
\end{pmatrix}.
$$
For fixed $(\delta_{\lambda},\delta_{\alpha^*},\delta_{\psi^*},\delta_\gamma)$, we can consider the behavior of $\delta_{\nu^*}$.
Let $\varphi_\tau(u)=\tau-I(u<0)$ and
$$
g_{it}(\delta_{\lambda},\delta_{\alpha^*},\delta_{\nu^*},\delta_{\psi^*},\delta_\gamma)=\frac{-1}{\sqrt{T}}\sum_{t=1}^{T}
\varphi_\tau\bigg(\varepsilon_{it}(\tau)-\frac{d_{it}\delta_\lambda}{\sqrt{NT}}-\frac{Z^*_{it}\delta_{\alpha^*}}{\sqrt{NT}}
-\frac{Z^*_{1i}\delta_{\nu^*}}{\sqrt{T}}-\frac{Z^*_{2t}\delta_{\psi^*}}
{\sqrt{N}}-\frac{\omega_{it}\delta_{\gamma}}{\sqrt{NT}}\bigg).
$$

Let
\begin{align*}
\quad\sup\|g_{it}(\delta_{\lambda},\delta_{\alpha^*},\delta_{\nu^*},\delta_{\psi^*},\delta_\gamma)-g_{it}(0,0,0,0,0)
\end{align*}
and
\begin{align*}
-\mathbb{E}[g_{it}(\delta_{\lambda},\delta_{\alpha^*},\delta_{\nu^*},\delta_{\psi^*},\delta_\gamma)
-g_{it}(0,0,0,0,0)]\|=o_p(1).
\end{align*}
Expanding $g_{it}$, we obtain
\begin{align*}
&\quad\mathbb{E}[g_{it}(\delta_{\lambda},\delta_{\alpha^*},\delta_{\nu^*},\delta_{\psi^*},\delta_\gamma)-g_{it}(0,0,0,0,0)] \\
&=\frac{-1}{\sqrt{T}}\sum_{t=1}^{T}\mathbb{E}\bigg(\varphi_\tau\bigg(\varepsilon_{it}(\tau)-\frac{d_{it}\delta_\lambda}{\sqrt{NT}}
-\frac{Z^*_{it}\delta_{\alpha^*}}{\sqrt{NT}}-\frac{Z^*_{1i}\delta_{\nu^*}}{\sqrt{T}} -\frac{Z^*_{2t}\delta_{\psi^*}}
{\sqrt{N}}-\frac{\omega_{it}\delta_{\gamma}}{\sqrt{NT}}\bigg)-\varphi_\tau(\varepsilon_{it}(\tau)\bigg)\\
&=\frac{-1}{\sqrt{T}}\sum_{t=1}^{T}\bigg[\tau-F\bigg(\xi_{it}(\tau)+\frac{d_{it}\delta_\lambda}{\sqrt{NT}}
+\frac{Z^*_{it}\delta_{\alpha^*}}{\sqrt{NT}}+\frac{Z^*_{1i}\delta_{\nu^*}}{\sqrt{T}} +\frac{Z^*_{2t}\delta_{\psi^*}}
{\sqrt{N}}+\frac{\omega_{it}\delta_{\gamma}}{\sqrt{NT}}\bigg)\bigg] \\
&=\frac{1}{\sqrt{T}}\sum_{t=1}^{T}f_{it}(\xi_{it}(\tau))\bigg[
\frac{d_{it}\delta_\lambda}{\sqrt{NT}}+\frac{Z^*_{it}\delta_{\alpha^*}}{\sqrt{NT}}+\frac{Z^*_{1i}\delta_{\nu^*}}{\sqrt{T}}
+\frac{Z^*_{2t}\delta_{\psi^*}}{\sqrt{N}}+\frac{\omega_{it}\delta_{\gamma}}{\sqrt{NT}}\bigg]+o_p(1),
\end{align*}
where $F(\cdot)$ is the conditional distribution of $y_{it}$. Obviously, $g_{it}(\hat\delta_{\lambda},\hat\delta_{\alpha^*},\hat\delta_{\nu^*},\hat\delta_{\psi^*},\hat\delta_\gamma)\rightarrow0$, and thus $\mathbb{E}[g_{it}(\delta_{\lambda},\delta_{\alpha^*},\delta_{\nu^*},\delta_{\psi^*},\delta_\gamma)-g_{it}(0,0,0,0,0)]=-g_{it}(0,0,0,0,0)$, i.e., the last equation has the following equivalent expression:
$$
\frac{1}{\sqrt{T}}\sum_{t=1}^{T}f_{it}(\xi_{it}(\tau))\bigg[\frac{d_{it}\delta_\lambda}{\sqrt{NT}}
+\frac{Z^*_{it}\delta_{\alpha^*}}{\sqrt{NT}}+\frac{Z^*_{1i}\delta_{\nu^*}}{\sqrt{T}} +\frac{Z^*_{2t}\delta_{\psi^*}}
{\sqrt{N}}+\frac{\omega_{it}\delta_{\gamma}}{\sqrt{NT}}\bigg]=
\frac{1}{\sqrt{T}}\sum_{t=1}^{T}\varphi_\tau(\varepsilon_{it}(\tau)).
$$
Optimality of $\hat\delta_{\nu^*}$ implies that $g_{it}(\delta_{\lambda},\delta_{\alpha^*},\delta_{\nu^*},\delta_{\psi^*},\delta_\gamma)=o(T^{-1})$, and thus
$$
\hat\delta_{\nu^*_i}=\bar{f}^{-1}_{i}\bigg(\frac{1}{\sqrt{T}}\sum_{t=1}^{T}\varphi_\tau(\varepsilon_{it}(\tau))
-\frac{1}{\sqrt{T}}\sum_{t=1}^{T}f_{it}(\xi_{it}(\tau))\bigg(\frac{d_{it}\delta_\lambda}{\sqrt{NT}}+\frac{Z^*_{it}\delta_{\alpha^*}}{\sqrt{NT}}+\frac{Z^*_{2t}\delta_{\psi^*}}{\sqrt{N}}
+\frac{\omega_{it}\delta_{\gamma}}{\sqrt{NT}}\bigg)\bigg)+o_p(1),
$$
where $\bar{f}_i=T^{-1}\sum_{t=1}^{T}f_{it}(\xi_{it}(\tau))$. Substituting $Z_{1i}^*\hat\delta_{\nu^*}$'s, we denote
$$
g_{t}(\delta_{\lambda},\delta_{\alpha^*},\delta_{\psi^*},\delta_\gamma)=\frac{-1}{\sqrt{N}}\sum_{i=1}^{N}
\varphi_\tau\bigg(\varepsilon_{it}(\tau)-\frac{d_{it}\delta_\lambda}{\sqrt{NT}}-\frac{Z^*_{it}\delta_{\alpha^*}}{\sqrt{NT}}
-\frac{Z^*_{1i}\hat\delta_{\nu^*}}{\sqrt{T}}-\frac{Z^*_{2t}\delta_{\psi^*}}
{\sqrt{N}}-\frac{\omega_{it}\delta_{\gamma}}{\sqrt{NT}}\bigg).
$$
Let
$$
\sup\|g_{t}(\delta_{\lambda},\delta_{\alpha^*},\delta_{\psi^*},\delta_\gamma)-g_{t}(0,0,0,0)
-\mathbb{E}[g_{t}(\delta_{\lambda},\delta_{\alpha^*},\delta_{\psi^*},\delta_\gamma)-g_{t}(0,0,0,0)]\|
=o_p(1).
$$
Expanding $g_{t}$, we obtain
\begin{align*}
&\quad\mathbb{E}[g_{t}(\delta_{\lambda},\delta_{\alpha^*},\delta_{\psi^*},\delta_\gamma)-g_{t}(0,0,0,0)] \\
&=\frac{-1}{\sqrt{N}}\sum_{i=1}^{N}\mathbb{E}\bigg(\varphi_\tau\bigg(\varepsilon_{it}(\tau)-\frac{d_{it}\delta_\lambda}{\sqrt{NT}}-\frac{Z^*_{it}\delta_{\alpha^*}}{\sqrt{NT}}
\frac{Z^*_{1i}\hat\delta_{\nu^*}}{\sqrt{T}}-\frac{Z^*_{2t}\delta_{\psi^*}}
{\sqrt{N}}-\frac{\omega_{it}\delta_{\gamma}}{\sqrt{NT}}\bigg)-\varphi_\tau(\varepsilon_{it}(\tau)\bigg)\\
&=\frac{1}{\sqrt{N}}\sum_{i=1}^{N}f_{it}(\xi_{it}(\tau))\bigg[\bigg(1-T^{-1}\bar{f}_i^{-1}\sum_{t=1}^{T}f_{it}(\xi_{it}(\tau))\bigg)
\bigg(\frac{d_{it}\delta_\lambda}{\sqrt{NT}}+\frac{Z^*_{it}\delta_{\alpha^*}}{\sqrt{NT}}
+\frac{Z^*_{2t}\delta_{\psi^*}}{\sqrt{N}}+\frac{\omega_{it}\delta_{\gamma}}{\sqrt{NT}}\bigg)\\
&\quad+T^{-1}\bar{f}_i^{-1}\sum_{t=1}^{T}\varphi_\tau(\varepsilon_{it}(\tau))\bigg]+o_p(1).
\end{align*}
Obviously, $g_{t}(\hat\delta_{\lambda},\hat\delta_{\alpha^*},\hat\delta_{\psi^*},\hat\delta_\gamma)\rightarrow0$, thus $\mathbb{E}[g_{t}(\delta_{\lambda},\delta_{\alpha^*},\delta_{\psi^*},\delta_\gamma)-g_{t}(0,0,0,0)]
=-g_{t}(0,0,0,0)$, i.e., the last equation has the following equivalent expression:
\begin{align*}
&\quad\frac{1}{\sqrt{N}}\sum_{i=1}^{N}f_{it}(\xi_{it}(\tau))\bigg[\bigg(1-T^{-1}\bar{f}_i^{-1}\sum_{t=1}^{T}f_{it}(\xi_{it}(\tau))\bigg)
\bigg(\frac{d_{it}\delta_\lambda}{\sqrt{NT}}+\frac{Z^*_{it}\delta_{\alpha^*}}{\sqrt{NT}}
+\frac{Z^*_{2t}\delta_{\psi^*}}{\sqrt{N}}+\frac{\omega_{it}\delta_{\gamma}}{\sqrt{NT}}\bigg)\\
&+T^{-1}\bar{f}_i^{-1}\sum_{t=1}^{T}
\varphi_\tau(\varepsilon_{it}(\tau))\bigg]=
\frac{1}{\sqrt{N}}\sum_{i=1}^{N}\varphi_\tau(\varepsilon_{it}(\tau)).
\end{align*}
Optimality of $\hat\delta_{\psi^*}$ implies that $g_{t}(\delta_{\lambda},\delta_{\alpha^*},\delta_{\psi^*},\delta_\gamma)=o(N^{-1})$, and thus
\begin{align*}
\hat\delta_{\psi^*_t}&=\bar{f}^{-1}_{t}\bigg(\frac{1}{\sqrt{N}}\bigg(1-T^{-1}\bar{f}_i^{-1}\sum_{t=1}^{T}f_{it}(\xi_{it}(\tau))\bigg)^{-1}\sum_{i=1}^{N}\varphi_\tau(\varepsilon_{it}(\tau))
-\sqrt{N}T^{-1}\bar{f}_t\bar{f}_i^{-1}\sum_{t=1}^{T}
\varphi_\tau(\varepsilon_{it}(\tau))\\
&\quad-\frac{1}{\sqrt{N}}\sum_{i=1}^{N}f_{it}(\xi_{it}(\tau))\bigg(\frac{d_{it}\delta_\lambda}{\sqrt{NT}}+\frac{Z^*_{it}\delta_{\alpha^*}}{\sqrt{NT}}
+\frac{\omega_{it}\delta_{\gamma}}{\sqrt{NT}}\bigg)\bigg)+o_p(1),
\end{align*}
where $\bar{f}_t=N^{-1}\sum_{i=1}^{N}f_{it}(\xi_{it}(\tau))$. Substituting $Z_{2t}^*\hat\delta_{\psi^*}$'s, we denote
$$
G(\delta_{\lambda},\delta_{\alpha^*},\delta_\gamma)=\frac{-1}{\sqrt{NT}}\sum_{i=1}^{N}\sum_{t=1}^{T}
\tilde{X}^T_{it}\varphi_\tau\bigg(\varepsilon_{it}(\tau)-\frac{d_{it}\delta_{\lambda}}{\sqrt{NT}}-\frac{Z^*_{it}\delta_{\alpha^*}}{\sqrt{NT}}-\frac{Z^*_{1i}\hat\delta_{\nu^*}}{\sqrt{T}}
-\frac{Z^*_{2t}\hat\delta_{\psi^*}}{\sqrt{N}}-\frac{\omega_{it}\delta_{\gamma}}{\sqrt{NT}}\bigg)
$$
where $\tilde{X}_{it}=[Z^*_{it},\omega_{it}]$. Let
$$
\sup\|G(\delta_{\lambda},\delta_{\alpha^*},\delta_\gamma)-G(0,0,0)
-\mathbb{E}[G(\delta_{\lambda},\delta_{\alpha^*},\delta_\gamma)-G(0,0,0)]\|=o_p(1).
$$
Expanding $G$, we obtain
\begin{align*}
&\quad\mathbb{E}[G(\delta_{\lambda},\delta_{\alpha^*},\delta_\gamma)-G(0,0,0)] \\
&=\frac{1}{\sqrt{NT}}\sum_{i=1}^{N}\sum_{t=1}^{T}
\tilde{X}^T_{it}f_{it}(\xi_{it}(\tau))\bigg[\frac{d_{it}\delta_{\lambda}}{\sqrt{NT}}+\frac{Z^*_{it}\delta_{\alpha^*}}{\sqrt{NT}}
+\frac{Z^*_{1i}\hat\delta_{\nu^*}}{\sqrt{T}}+\frac{Z^*_{2t}\hat\delta_{\psi^*}}{\sqrt{N}}
+\frac{\omega_{it}\delta_{\gamma}}{\sqrt{NT}}\bigg]+o_p(1), \\
&=\frac{1}{\sqrt{NT}}\sum_{i=1}^{N}\sum_{t=1}^{T}
\tilde{X}^T_{it}f_{it}(\xi_{it}(\tau))\bigg[\bigg(1-N^{-1}\bar{f}_t^{-1}\sum_{i=1}^{N}f_{it}(\xi_{it}(\tau))\bigg)
\bigg(1-T^{-1}\bar{f}_i^{-1}\sum_{t=1}^{T}f_{it}(\xi_{it}(\tau))\bigg) \\
&\quad\bigg(\frac{d_{it}\delta_{\lambda}}{\sqrt{NT}}+\frac{Z^*_{it}\delta_{\alpha^*}}{\sqrt{NT}}
+\frac{\omega_{it}\delta_{\gamma}}{\sqrt{NT}}\bigg)+N^{-1}\bar{f}_t^{-1}\sum_{i=1}^{N}\varphi_\tau(\varepsilon_{it}(\tau))
+T^{-1}\bar{f}_i^{-1}\sum_{t=1}^{T}\varphi_\tau(\varepsilon_{it}(\tau))\bigg]
+o_p(1).
\end{align*}
Obviously, $G(\hat\delta_{\lambda},\hat\delta_{\alpha^*},\hat\delta_\gamma)\rightarrow0$, $\mathbb{E}[G(\delta_{\lambda},\delta_{\alpha^*},\delta_\gamma)-G(0,0,0)]=-G(0,0,0)$, i.e., the last equation has the following equivalent expression:
\begin{align*}
&\quad\frac{1}{\sqrt{NT}}\sum_{i=1}^{N}\sum_{t=1}^T\tilde{X}^T_{it}\varphi_\tau(\varepsilon_{it}(\tau))=
\frac{1}{\sqrt{NT}}\sum_{i=1}^{N}\sum_{t=1}^{T}
\tilde{X}^T_{it}f_{it}(\xi_{it}(\tau))\bigg[\bigg(1-N^{-1}\bar{f}_t^{-1}\sum_{i=1}^{N}f_{it}(\xi_{it}(\tau))\bigg)\\
&\bigg(1-T^{-1}\bar{f}_i^{-1}\sum_{t=1}^{T}f_{it}(\xi_{it}(\tau))\bigg)
\bigg(\frac{d_{it}\delta_{\lambda}}{\sqrt{NT}}+\frac{Z^*_{it}\delta_{\alpha^*}}{\sqrt{NT}}
+\frac{\omega_{it}\delta_{\gamma}}{\sqrt{NT}}\bigg)+N^{-1}\bar{f}_t^{-1}\sum_{i=1}^{N}\varphi_\tau(\varepsilon_{it}(\tau)) \\
&+T^{-1}\bar{f}_i^{-1}\sum_{t=1}^{T}\varphi_\tau(\varepsilon_{it}(\tau))\bigg].
\end{align*}
Letting $\delta_{\zeta}=(\alpha^T_{\alpha^*},\alpha^T_{\gamma})^T$, we  write the equation above as:
\begin{align*}
&\quad\frac{1}{\sqrt{NT}}\sum_{i=1}^{N}\sum_{t=1}^{T}
\tilde{X}^T_{it}f_{it}(\xi_{it}(\tau))\bigg[\bigg(1-N^{-1}\bar{f}_t^{-1}\sum_{i=1}^{N}f_{it}(\xi_{it}(\tau))\bigg)\\
&\bigg(1-T^{-1}\bar{f}_i^{-1}\sum_{t=1}^{T}f_{it}(\xi_{it}(\tau))\bigg)
\bigg(\frac{\tilde{X}_{it}\delta_{\zeta}}{\sqrt{NT}}+\frac{d_{it}\delta_{\lambda}}
{\sqrt{NT}}\bigg)+N^{-1}\bar{f}_t^{-1}\sum_{i=1}^{N}\varphi_\tau(\varepsilon_{it}(\tau))\\
&+T^{-1}\bar{f}_i^{-1}\sum_{t=1}^{T}\varphi_\tau(\varepsilon_{it}(\tau))\bigg]
=\frac{1}{\sqrt{NT}}\sum_{i=1}^{N}\sum_{t=1}^T\tilde{X}^T_{it}\varphi_\tau(\varepsilon_{it}(\tau)).
\end{align*}

Alternatively, using more convenient notation, we write the last expression as:
$$
\mathbf{J}_{\zeta}\delta_{\zeta}+\mathbf{J}_{\lambda}\delta_{\lambda}=\mathbb{J}_{\phi},
$$
where $\mathbf{J}_{\zeta}=\underset{N,T\rightarrow\infty}{\lim}\tilde{X}^{T}M^T_{\tilde{Z}}\Omega M_{\tilde{Z}}\tilde{X}$, $\mathbf{J}_{\lambda}=\underset{N,T\rightarrow\infty}{\lim}\tilde{X}^{T}M^T_{\tilde{Z}}\Omega M_{\tilde{Z}}D$, $\mathbb{J}_{\phi}$ is a mean zero r.v. with
covariance $\tau(1-\tau)\tilde{X}^TM^T_{\tilde{Z}}M_{\tilde{Z}}\tilde{X}$, $\Omega=diag(f_{it}(\xi_{it}(\tau)))$ and $\Phi_{\tau}$ is a $NT$-vector $(\phi_\tau(\varepsilon_{it}(\tau)))$, $\tilde{Z}=[Z_1^*,Z_2^*]$, $M_{\tilde{Z}}=I-P_{\tilde{Z}}$, $P_{\tilde{Z}}=\tilde{Z}(\tilde{Z}^T\Omega\tilde{Z})^{-1}\tilde{Z}^T\Omega$.

Letting $[\bar{\mathbf{J}}^T_{\alpha^*},\bar{\mathbf{J}}_\gamma^T]$ be a
conformable partition of $\mathbf{J}^{-1}_{\zeta}$ as in Galvao (2011) and Chernozhukov and Hansen (2006) yields
$\hat\delta_{\alpha^*}=\bar{\mathbf{J}}^T_{\alpha^*}(\mathbb{J}_\phi-\mathbf{J}_{\lambda}\delta_{\lambda})$,
and $\hat\delta_{\gamma}=\bar{\mathbf{J}}^T_{\gamma}(\mathbb{J}_\phi-\mathbf{J}_{\lambda}\delta_{\lambda})$.
Letting $H=\bar{\mathbf{J}}^T_\gamma A\bar{\mathbf{J}}_\gamma$ as in Chernozhukov and Hansen (2006) gives
$\hat\delta_{\lambda}=K\mathbb{J}_\phi$, where $K=(\mathbf{J}_{\lambda}^{T}H\mathbf{J}_{\lambda})^{-1}\mathbf{J}_{\lambda}^{T}H$.
Replacing it in the previous expression, $\hat\delta_{\gamma}=\bar{\mathbf{J}}^T_{\gamma}(\mathbb{J}_\phi-\mathbf{J}_{\lambda}\delta_{\lambda})
=\bar{\mathbf{J}}^T_{\gamma}(I-\mathbf{J}_{\lambda}(\mathbf{J}_{\lambda}^{T}H\mathbf{J}_{\lambda})^{-1}
\mathbf{J}_{\lambda}^{T}H)\mathbb{J}_\phi=\bar{\mathbf{J}}^T_{\gamma}M
\mathbb{J}_\phi$, where $M=I-\mathbf{J}_{\lambda}(\mathbf{J}_{\lambda}^{T}H\mathbf{J}_{\lambda})^{-1}
\mathbf{J}_{\lambda}^{T}H$. Due to the  invertibility of $\mathbf{J}_{\lambda}\bar{\mathbf{J}}_{\gamma}$,
$\hat\delta_\gamma=\mathbf{0}\times O_p(1)+o_p(1)$. Similarly, substituting back $\delta_{\lambda}$, we obtain that $\hat\delta_{\alpha^*}=\bar{\mathbf{J}}^T_{\alpha^*}M\mathbb{J}_\phi$. By the regularity conditions, we have that
$$
\begin{pmatrix}
\hat\delta_{\lambda}(\lambda_n,\tau)\\
\hat\delta_{\alpha^*}(\lambda_n,\tau)
\end{pmatrix}=
\begin{pmatrix}
\sqrt{NT}(\hat{\lambda}(\lambda_n,\tau)-\lambda(\tau))\\
\sqrt{NT}(\hat{\alpha}^*(\lambda_n,\tau)-\alpha^*(\tau))
\end{pmatrix}\rightsquigarrow\mathcal{N}(\mathbf{0},J^TSJ).
$$

\end{document}